\documentstyle[12pt]{article}
\newcommand{\slq}{\raise.15ex\hbox{$/$}\kern-.57em\hbox{$q$}}
\newcommand{\slp}{\raise.15ex\hbox{$/$}\kern-.57em\hbox{$p$}}
\newcommand{\be}{\begin{equation}}
\newcommand{\ee}{\end{equation}}
\newcommand{\bear}{\begin{eqnarray}}
\newcommand{\ear}{\end{eqnarray}}

\date{}
\begin{document}
\begin{titlepage}
\begin{flushright}
HD--THEP--97--5
\end{flushright}
\quad\\
\vspace{1.8cm}
\begin{center}
{\bf \LARGE Quantum Dynamics}\\
\medskip
{\bf\LARGE in Classical Time Evolution}\\
\medskip
{\bf\LARGE of Correlation Functions}\\
\vspace{1cm}
Christof Wetterich\footnote{This work was performed in part
at ITP, UCSB, Santa Barbara, and supported by
the National Science Foundation under Grant No. PHY94-07 197.}\\
\bigskip
Institut  f\"ur Theoretische Physik\\
Universit\"at Heidelberg\\
Philosophenweg 16, D-69120 Heidelberg\\
C.Wetterich@thphys.uni-heidelberg.de
\vspace{1cm}
\begin{abstract}
The time-dependence of correlation functions
under the influence of classical equations of motion
is described by an exact evolution equation.
For conservative systems thermodynamic equilibrium is a fixed
point of these equations. We show that this fixed point is not
universally stable, since infinitely many conserved correlation
functions obstruct the approach to equilibrium.
Equilibrium can therefore be reached at most for suitably
averaged quantities or for subsystems, similar to quantum statistics.
The classical time evolution of correlation functions
shows many dynamical features of quantum mechanics.
\end{abstract}
\end{center}
\end{titlepage}
\newpage
Recent years have seen large-scale computer simulations of
the statistical behaviour of nonlinear classical field equations.
Examples in particle physics and cosmology range from the formation
of defects during phase transitions in the early universe
and inflation \cite{1} to the computation of the rate of baryon-number
violating processes at high temperature $T$ \cite{2}
or speculations
about the formation of a disordered chiral condensate in
heavy ion collisions \cite{3}. A typical model is a
$\varphi^4$ theory for $N$ complex scalar fields,
$N=1$ being relevant for string formation and $N=2$ for the chiral
phase transition. Simulations of statistical properties of such
systems solve the classical nonlinear field equations numerically for
given initial conditions. Subsequently the average over a suitable
ensemble of initial conditions is taken. This determines
the time-dependent correlation functions of the system. The use
of classical field equations is often motivated by the conjecture
 that classical equations govern
the behaviour of long-distance
modes (typical momenta smaller than $T$) even in case of
high temperature quantum field theory. We emphasize that quantum
effects can be (partially) incorporated by using the field equations
derived from the quantum-effective action or a coarse-grained
version of it \cite{4}.

As a step towards an analytical treatment of the time evolution
of statistical properties an exact evolution equation for generating
functionals of time-dependent (``equal time''-) correlation
functions has recently been proposed \cite{5}.
For conservative classical field equations
thermodynamic equilibrium is a fixed point
of this evolution equation. In this note we investigate
if and how this fixed point is approached. Intuitively, one
expects that suitably averaged ``macroscopic'' quantities
should thermalize even in the absence of dissipation:
Two types of gas should mix due to classical motion with
elastic scattering of the molecules.\footnote{In case of a finite
number of conserved quantities beyond energy the equilibrium
should, of course, be formulated in
accordance with the conservation laws -- in analogy to a conserved
particle number in classical statistics.} Nevertheless, our
investigations show that the equilibrium fixed point is
not universally attractive. We find that fluctuations around
this fixed point are characterized by infinitely many conserved
correlations which obstruct the approach to the fixed point. The
situation seems very similar to quantum statistics where the
trace of arbitrary powers of the density matrix is conserved
for a closed system. Under such cirucumstances only suitable
``macroscopic'' degrees of freedom or particular subsystems
can thermalize whereas on a ``microscopic'' scale correlation
functions fluctuate in time. Our treatment of the time evolution
of correlation functions in classical statistics indeed
reveals many analogies to quantum statistics. Our considerations
are quite general, with applications
far beyond the specific $\varphi^4$-models mentioned in this note.

Consider some number of complex degrees of freedom $\chi_m$ whose time
evolution is given by the equation of motion
\be\label{1}
\frac{d}{dt}\chi_m\equiv\dot\chi_m=F_m[\chi]
=-i\hat H_{mn}[\chi]\chi_n\ee
(Summation over repeated indices is always implied.)
The matrix $\hat H$ is hermitean and we assume that it can
be expanded in powers of $\chi$. For the purpose of this note
we will also specialize to the case of a conserved charge
(corresponding to phase rotations of $\chi$) for which
$\hat H$ contains equal powers of $\chi$ and $\chi^*$
\be\label{2}
\hat H_{mn}=\sum_{l=0}^\infty\hat h^{(l)}_{mnq_1r_1...q_lr_l}
\chi^*_{q_1}\chi_{r_1}...\chi^*_{q_l}\chi_{r_l}\ee
Here $\hat h^{(l)}$ is symmetric in $(m,q_1,...,q_l)$
and $(n,r_1,...,r_l)$ and obeys $\hat h^{(l)*}_{mnq_1r_1...
q_lr_l}=\hat h^{(l)}_{nmr_1q_1...r_lq_l}$.
The equation of motion (\ref{1}) can be derived
from a quantity $U$
\be\label{3}
U=\sum_{l=0}^\infty\frac{1}{l+1}\hat h^{(l)}_{mnq_1r_1...q_lr_l}
\chi^*_m\chi_n\chi^*_{q_1}\chi_{r_1}...\chi^*_{q_l}\chi_{r_l}\ee
as
\be\label{4}
\dot\chi_m=-i\frac{\partial U}{\partial\chi_m^*},\quad
\dot\chi_m^*=i\frac{\partial U}{\partial\chi_m}\ee
We note that the system is conservative $(dU/dt=0)$
if the coefficients $\hat h^{(l)}$ are time-independent.
Eqs. (\ref{3}), (\ref{4}) are characteristic for non-relativistic
field equations.
As an example we start from the relativistic
wave equation for a complex scalar field $\varphi$
\be\label{5}
\ddot\varphi(x)=(\Delta-M^2)\varphi(x)-\frac{\partial W}
{\partial\rho}(x)\varphi(x)\ee
Here $W$ is a real function of $\rho=\varphi^*\varphi$.
In the nonrelativistic limit we use
\be\label{6}
\varphi(x,t)=e^{-iMt}\chi(x,t)\ee
and neglect terms $\sim \ddot\chi$ such that
\be\label{7}
\dot\chi(x)=-i\{-\frac{\Delta}{2M}+\frac{1}{2M}
\frac{\partial W}{\partial\rho}(x)\}\chi(x)\ee
We identify the infinitely many $\chi_m$ with $\chi(x)$ or with
corresponding Fourier modes $\chi(p)=\int d^dxe^{-ip_ix_i}\chi(x)$.
Obviously $\hat H=-\frac{\Delta}{2M}+\frac{1}{2M}\frac{\partial
W}{\partial\rho}(x)$ is hermitean. We have used a notation
where eq. (\ref{1}) ressembles the Schr\"odinger equation.
It should be emphasized , however, that in our case $\hat H$ is not
a linear operator unless the equation of motion is linear
($\hat h^{(l)}=0$ for $l>0$). In the following we will
also include the possibility that the coefficients $\hat h
^{(l)}$ depend on time.
We will see below that crucial features of the non-relativistic
limit also persist for the relativistic second-order equation
(\ref{5}).

At some time $t_0$ we specify an ensemble of initial
conditions for $\chi^0_m=\chi_m(t_0)$ by a probability
distribution $\exp-S_0[\chi^0]$.
At any later time $t$ the generating functional for the
$n$-point functions is given by
\be\label{8}
Z[j,t]=\int D\chi^0\exp\{-S_0[\chi^0]+j^*_m\chi_m(t;\chi^0)
+j_m\chi_m^*(t;\chi^0)\}\ee
with  $\chi_m(t;\chi^0)$ the solution of the equation
of motion (\ref{1}) with the particular initial
condition $\chi(t_0)=\chi^0$.
In particular, the two-point function
reads
\be\label{9}
\rho_{mn}(t)=<\chi_m(t)\chi^*_n(t)>=Z^{-1}\frac{\partial^2Z}
{\partial j^*_m\partial j_n}
|_{j=0}\ee
with $Z=Z[j=0]$ independent of $t$. More generally, for a given time
$t$ the state of the system can be characterized by the
coefficients $z^{(k,l)}(t)$ of a Taylor expansion of
$Z[j,t]$
\be\label{10}
Z[j,t]=\sum^\infty_{k=0}\sum^\infty_{l=0}\frac{1}{k!l!}
z_{q_1...q_kr_1...r_l}^{(k,l)}(t)j^*_{q_1}...j^*_{q_k}
j_{r_1}...j_{r_l}\ee
with
\be\label{11}
<\chi_{q_1}(t)...\chi_{q_k}(t)\chi^*_{r_1}(t)...\chi^*_{r_l}(t)>
=Z^{-1}z^{(k,l)}_{q_1...q_kr_1...r_l}(t)\ee
The time dependence of $Z[j,t]$ is determined
by an exact evolution equation \cite{Wet1}
\bear\label{12}
\partial_tZ[j]&=&j^*_mF_m[\frac{\partial}{\partial j}
]Z[j]+j_mF_m^*[\frac{\partial}{\partial j}]Z[j]\nonumber\\
&=&-ij^*_m\sum^\infty_{l=0}\hat h^{(l)}_{mnq_1r_1...q_lr_l}
\frac{\partial}{\partial j^*_n}\frac{\partial}{\partial j_{q_1}}
\frac{\partial}{\partial j^*_{r_1}}...\frac{\partial}{\partial j_{q_l}}
\frac{\partial}{\partial j^*_{r_l}}Z[j]\nonumber\\
&&+ij_m\sum^\infty_{l=0}\hat h^{(l)}_{nmq_1r_1...q_lr_l}
\frac{\partial}{\partial j_n}\frac{\partial}{\partial j_{q_1}}
\frac{\partial}{\partial j^*_{r_1}}...\frac{\partial}{\partial j_{q_l}}
\frac{\partial}{\partial j^*_{r_l}}Z[j]\ear
with $\partial_t$ the time derivative at fixed $j$.
This is equivalent to an infinite system of
evolution equations for the correlation functions $z^{(k,l)}$,
which obtains by taking appropriate partial derivatives
of eq. (\ref{12}) with respect to $j$. For the equation of
motion (\ref{1}) $F_m$ involves at least one derivative
with respect to $j$. In consequence, the evolution equation for
$z^{(k,l)}$ involves the correlation functions
$z^{(k',l')}$ with $k'+l'\geq k+l$, but not those with
$k'+l'<k+l$. For a given $k+l$ the correlation functions
$z^{(k',l')}$ with $k'+l'>k+l$ act as a sort of ``environment''
for the evolution of $z^{(k,l)}$: They influence the dynamics
of $z^{(k,l)}$, but are not affected themselves by the time
evolution of $z^{(k,l)}$. Symmetries of $Z[j]$ are conserved
\footnote{More precisely, this holds only for symmetry transformations
that do not mix $j_m$ with $j^*_m$. The transformations of $j_m$
and $\chi_m$ are related by the invariance of $j^*_m\chi_m+j_m\chi^*
_m$.}
by the evolution equation (\ref{12}) if $U$ is invariant
under these symmetries. If $Z$ involves initially only terms
with an equal number of $j_m$ and $j^*_m$, this property
remains conserved by the evolution. More generally, the
correlation functions $z^{(k,l)}$ can be classified by
their charge $Q=k-l$. Sectors with different charge are
not mixed by the time evolution. If
$S_0[\chi^0]$ in eq. (\ref{8}) is neutral, only the neutral sector of
$Z[j]$ is nonvanishing.

For conservative equations of motion (\ref{4}), eq. (\ref{12})
has a fixed point $(\partial_tZ_*=0)$ which corresponds to
thermodynamic equilibrium at temperature $T=\beta^{-1}$
\cite{5}, namely
\be\label{13}
Z_*[j,\beta]=\int D\chi\exp(-\beta U[\chi]+j^*_m\chi_m+j_m\chi
^*_m)\ee
In this note we address the question if this fixed point
is approached for $t\to\infty$. We will see that for generic
initial conditions this is not the case due to the existence
of infinitely many conserved quantities which
characterize the evolution of the correlation functions. For
the equation of motion (\ref{1}) a first set of
conserved quantities is easily found by observing that due to
the factor $i$ one has for arbitrary $j$
\bear\label{14}
\partial_t\frac{\partial^2Z}{\partial j^*_m\partial j_m}
&=&\frac{\partial^2}{\partial j^*_m\partial j_m}(j^*_p
F_p[\frac{\partial}{\partial j}]
+j_pF^*_p[\frac{\partial}{\partial j}])Z[j]\nonumber\\
&=&(j_p^*F_p[\frac{\partial}{\partial j}]+j_pF^*_p
[\frac{\partial}{\partial j}])\frac{\partial^2}
{\partial j_m^*\partial j_m}Z[j]\ear
For $j=0$ one infers immediately that the trace of the two-point
function (\ref{9}) is conserved, $dTr\rho/dt=0$. We also
conclude that the operators $(\partial^2/\partial j^*_m\partial j_m)$
and $(j_p^*F_p[\frac{\partial}{\partial j}]+j_pF^*_p
[\frac{\partial}{\partial j}])$ commute. This leads to an
infinite set of conserved correlation functions $X_N$
\be\label{15}
\frac{dX_N}{dt}=\frac{d}{dt}\left\{Z^{-1}
\left(\frac{\partial^2}{\partial j^*_m\partial j_m}\right)
^NZ\right\}_{j=0}=0\ee
It follows immediately that the equilibrium fixed point (\ref{13})
cannot be reached unless the initial values of \underbar{all}
$X_N$ exactly correspond to the ones of the
equilibrium distribution!

Actually, the $X_N$ are only a small subset of the conserved
correlation functions. We will concentrate in the following
on the two-point function $\rho_{mn}$ and show that
there are states for which this will behave exactly
as the density matrix in quantum statistics. The evolution
equation for the two-point function reads
\be\label{16}
Z\partial_t\rho_{mn}=-i\hat H_{mp}[\frac{\partial}{\partial j}]
\frac{\partial^2Z}{\partial j_p^*\partial j_n}_{|j=0}\nonumber\\
+i\hat H_{pm}[\frac{\partial}{\partial j}]\frac{\partial^2Z}
{\partial j_m^*\partial j_p}_{|j=0}\ee
We note that only those terms in the general expansion
of $Z[j]$ (\ref{10})
which have zero charge
\be\label{17}
Z[j]=\sum^{\infty}_{l=0}\frac{1}{(l!)^2}
\hat z^{(l)}_{q_1r_1...q_lr_l}j^*_{q_1}j_{r_1}...j^*_{q_l}j_{r_l}+...\ee
contribute to eq. (\ref{16})
\bear\label{18}
\frac{d}{dt}\rho_{mn}&=&-iY_{mn}+iY^\dagger_{mn}\nonumber\\
Y_{mn}&=&Z^{-1}\hat H_{mp}[\frac{\partial}{\partial j}]
\frac{\partial^2Z}{\partial j^*_p\partial j_n}_{|j=0}
\nonumber\\
&&=\hat h^{(0)}_{mp}\rho_{pn}+Z^{-1}\sum^\infty_{l=1}\hat h
^{(l)}_{mpq_1r_1...q_lr_l}\hat z^{(l+1)}_{pnr_1q_1...r_lq_l}\ear
We will consider states which obey
\be\label{19}
Y-Y^\dagger=[H,\rho]-i\{K,\rho\},\quad
H=H^\dagger,\ K=K^\dagger\ee
This holds, in particular, whenever $\rho$ is invertible,
$H-iK=Y\rho^{-1}$.
In terms of the matrices $H$ and $K$ we can write
\be\label{20}
\frac{d}{dt}\rho=-i[H,\rho]-\{K,\rho\}\ee
with $Tr(K\rho)=0$.
The particular subclass of states which, in addition to the
condition (\ref{19}), obeys for all $t$
\be\label{21}
\{K,\rho\}=0\ee
will be called \underbar{quantum states}. The time evolution of
quantum states is governed by a von Neumann-type equation
$d\rho/dt=-i[H,\rho]$.
In particular, for linear equations of motion
only $\hat h^{(0)}$ contributes to $Y$ (\ref{18})
and all states are quantum states. For nonlinear equations
of motion the Hamiltonian $H$ still depends on the particular
quantum state which is not only characterized by $\rho$, but
also by higher correlation functions $\hat z^{(l)}$.
Nevertheless, the commutator on the r.h.s.
implies that the trace of arbitrary powers of $\rho$
is conserved
\be\label{22}
\frac{d}{dt}Tr(\rho^N)=0\ee
This yields once more an infinite set of conserved quantities.
Relation (\ref{22}) is well known for the density matrix
in quantum mechanics. For any function $F(\rho)$ which admits
a Taylor expansion it implies $\frac{d}{dt}Tr\ F(\rho)=0$.
In particular, the quantum mechanical entropy $S_Q=-Tr\ \rho
\ \ln\ \rho$ is conserved. \underline{Linear quantum states} are
those for which $H$ does not depend on $\rho$.

The analogy of $\rho$ with the quantum mechanical density matrix
can be pushed even further. By the defintion (\ref{9})
$\rho$ is hermitean and obeys the positivity conditions
$\xi_m\rho_{mn}\xi^*_n\geq0$
for arbitrary complex vectors $\xi$. In particular, the diagonal
elements of $\rho$ are positive semidefinite, $\rho_{nn}\geq0$, and obey
$\rho_{mm}\rho_{nn}\geq|\rho_{mn}|^2$ for all $m$ and $n$.
The expectation value of an arbitrary quantity which is bilinear
in $\chi$, ${\cal A}=\chi_m^*A_{mn}\chi_n$,
can be written in terms of an operator associated to the matrix
$A_{mn}$ as $<{\cal A}>=Tr(A\rho)$.
Different operators $A, B$ do not necessarily commute.
By suitable unitary transformations one can bring $\rho$
into diagonal form $\rho=diag(p_n)$ with
real positive semidefinite eigenvalues $p_n\geq0$. Therefore
$\rho$ is invertible if none of the $p_n$
vanishes.

In order to gain some intuition how restrictive
is the ``quantum condition'' (\ref{21}) we consider
an equation of motion which contains only quartic nonlinearities,
i.e. $\hat h^{(l)}=0$ for $l\geq2$. For invertible $\rho$
the conditon $\{K,\rho\}=0$
for a quantum state is equivalent to $K=0$ or
$\rho Y=Y^\dagger\rho$, as can be seen easily in a basis
where $\rho$ is diagonal. Then quantum
states obey a condition for the four-point function $\hat z^{(2)}$
\be\label{23}
\hat h^{(1)}_{spqr}(\rho_{ms}\hat z^{(2)}_{pnrq}-
\rho_{pn}\hat z^{(2)}_{smrq})=0\ee
Since this condition should hold for all $t$, the time
derivative of the l.h.s. must also vanish. In turn, this
gives a condition for $\hat z^{(3)}$, and so forth for higher
time derivatives. Generically, quantum states therefore correspond
to restrictions for all $\hat z^{(l)}$. Nevertheless,
$\hat z^{(2)}_{pnqr}$ has four indices and eq. (\ref{23})
is an equation for a combination forming a matrix with two indices.
One therefore expects that there are many solutions of eq. (\ref{23})
for $\hat z^{(2)}$ with given $\hat h$ and
$\rho$. The occurrence of quantum states obeying the conditions
(\ref{19}) and (\ref{21}) seems quite generic!\footnote{Mathematically,
the simplest quantum state is $\hat z^{(l)}=0$ for $l\geq2$. This
seems, however, not a very natural state for physical systems.}

As an example for our discussion of quantum states, let us take the
field equation (\ref{7}) with a local quartic interaction
$W(\rho)=\frac{1}{2}\lambda\rho^2$. In a basis
 where the index $m$ denotes space coordinates (i.e. lattice sites),
the nonlinear part of $\hat H$ takes the form $\hat h^{(1)}_{mnqr}=
\frac{\lambda}{4M}\delta_{mq}(\delta_{mn}
\delta_{mr}+\delta_{qn}\delta_{qr})$. The evolution equation for the
two-point function reads
\bear\label{24}
&&\frac{d}{dt}\rho(x,y)=\frac{d}{dt}<\chi(x)\chi^*(y)>
=-i\Bigl\{-\frac{1}{2M}(\Delta_x-\Delta_y)\rho(x,y)\nonumber\\
&&+\frac{\lambda}{2M}(<\chi(x)\chi^*(x)\chi(x)\chi^*(y)>
-<\chi(x)\chi^*(y)\chi(y)\chi^*(y)>)\Bigr\}\ear
The conditions for a quantum state require that for every $t$
there exists a function $G(x,z)=G^*(z,x)$ such that
\bear\label{25}
&&<\chi(x)\chi^*(y)\chi(x)\chi^*(x)>-
<\chi(x)\chi^*(y)\chi(y)\chi^*(y)>\nonumber\\
&&=\int dz\{G(x,z)<\chi(z)\chi^*(y)>-G(z,y)<\chi(x)\chi^*
(z)>\}\ear
We will next consider a particular class of states where all
correlation functions $\hat z^{(l)}$ are charge
conjugation- and translation-invariant for $l\geq2$.
For these states the ``quantum condition'' (\ref{25}) will
be fulfilled by virtue of the symmetries with
$G(x,z)=0$. Since $U=\frac{1}{2M}\int d^dx\{\partial_i
\chi^*(x)\partial_i\chi(x)+\frac{\lambda}{2}(\chi^*(x)\chi(x))^2\}$
is translation-invariant, an initial translation invariance
of the correlation functions $\hat z^{(l)}$ for $l\geq 2$ is
preserved by the evolution. Due to the particular structure
of the evolution equation (\ref{12}),
this property even holds if $\hat z^{(1)}$ is not translation-invariant.
(A translation-invariant ``environment'' for $\hat z^{(1)}$ is
not affected by the evolution of $\hat z^{(1)}$.) In a momentum basis
the translation invariance of $\hat z^{(2)}$ implies
\be\label{26}
<\chi(p_1)\chi^*(p_2)\chi(p_3)\chi^*(p_4)>=
A_4(p_1,p_2,p_3,p_4)(2\pi)^d\delta(p_1-p_2+p_3
-p_4)\ee
and the evolution equation for $\rho(p,q)=<\chi(p)\chi^*(q)>$
reads
\bear\label{27}
&&\frac{d}{dt}\rho(p,q)=-\frac{i}{2M}(p^2-q^2)\rho(p,q)\\
&&-\frac{i\lambda}{2M}\int\frac{d^dr}{(2\pi)^d}\frac{d^ds}{(2\pi)^d}
[A_4(r,p,s,r+s-p)
-A_4(p,r,r+s-p,s)](2\pi)^d\delta(p-q)\nonumber\ear
In addition, $U$ is invariant under the charge conjugation
$\chi(p)\to\chi^*(p)$. For charge conjugation symmetric states
one has $A_4(p_1,p_2,p_3,p_4)=A_4(p_2,p_1,p_4,p_3)$ and the
nonlinear term $\sim\lambda$ in eq. (\ref{27}) vanishes.
For a translation- and charge conjugation-invariant
``environment'' the evolution of the ``density matrix''
$\rho$ is therefore governed by a linear von Neumann equation
\be\label{28}
\frac{d}{dt}\rho(p,q)=-i\int\frac{d^ds}{(2\pi)^d}(H(p,s)\rho
(s,q)-\rho(p,s)H(s,q))\ee
with a hermitean free Hamiltonian
$H(p,q)=\frac{p^2}{2M}(2\pi)^d\delta(p-q)$.
Despite the nonlinearity of the microscopic equation of motion
(\ref{7}) we have found a class of probability distributions
for which the two-point function obeys a \underline{linear}
evolution equation! We should emphasize, however, that charge
conjugation invariance of the initial distribution does
not guarantee that the state (or the environment for
$\hat z^{(1)}$) remains invariant at later times. Charge conjugation
maps $\chi$ onto $\chi^*$ and is therefore not automatically
conserved by the evolution equation. In general, only a subclass
of states preserves charge conjugation symmetry. In particular,
the thermal equilibrium states (\ref{13}) are translation- and charge
conjugation-invariant. If we consider deviations from these equilibrium
states where only the two-point function $\hat z^{(1)}$
differs from the equilibrium value, the time evolution of $\rho$
is governed by the von Neumann equation (\ref{26}).
The superposition principle is valid, and for two solutions
$\rho_1$ and $\rho_2$ the sum $\rho_1+\rho_2$ is also a solution.

The construction of the general solution of the von Neumann
equation (with arbitrary hermitean $H$) is well known from
quantum mechanics: It is based on a set of wave functions
$\psi^{(n)}(t)$ which obey the Schr\"odinger equation
\be\label{29}
\frac{d}{dt}\psi^{(n)}(t)=-iH\psi^{(n)}(t)\ee
with
\be\label{30}
\rho(t)=\sum_np_n\psi^{(n)}(t)\otimes\psi^{(n)*}(t)\ee
Here $p_n$ are the positive semidefinite time-independent
eigenvalues of $\rho$ and $\psi^{(n)}(t)$ the associated
eigenvectors in a standard normalization $\psi_s^{(m)*}\psi_s
^{(n)}=\delta^{mn}$. (In a given basis for the degrees of
freedom $\chi_m$ we can write $\psi^{(n)}
(t)=a_{nm}(t)\tilde\psi_m$ with time-independent basis
vectors $\tilde\psi_m$ such that $\rho_{mn}(t)=\sum_sp_s
a_{sm}a^*_{sn}$.) A \underline{pure} linear quantum state
obeys $\rho^2=\rho\ Tr\ \rho$. So far we have not chosen a
particular normalization of $\rho$ -- the standard normalization
$Tr\ \rho=1$ can always be achieved by a suitable rescaling
of the degrees of freedom $\chi_m$.
In summary, we have found a class of states for which
the two-point function obeys
all the dynamical rules of a quantum-mechanical
density matrix. The time evolution
of such states will not approach the thermodynamic equilibrium state.
Just like in quantum mechanics, thermodynamic equilibrium
can only be attained for suitable subsystems or macroscopic
degrees of freedom.

For the non-relativistic quartic scalar theory, we find from eq. (\ref{28})
that a translation symmetric two-point function
\be\label{31}
\rho(p,q)=A_2(p)(2\pi)^d\delta(p-q)\ee
is conserved for arbitrary (positive semidefinite)
$A_2(p)$ if the higher $n$-point functions are the ones of a
linear quantum state. We have therefore found an infinity of
new fixed points of the evolution equation (\ref{12}). They are related
to the thermal equilibrium state (\ref{13}) by
\be\label{32}
Z_*^{(A_2)}[j]=Z_*[j,\beta]
+Z\int\frac{d^dp}{(2\pi)^d}j^*(p)(A_2(p)-A_{2*}^{(\beta)}(p))j(p)\ee
with $A_{2*}^{(\beta)}(p)$ the two-point function in thermal
equilibrium. We emphasize that the connected higher $n$-point
functions generated by $W_*^{(A_2)}[j]=\ln\ Z_*^{(A_2)}[j]$
differ from the ones in thermal equilibrium. (The disconnected
parts are modified by the difference between $A_2$ and $A_{2*}^{(\beta)}$
which implies that the connected parts must also be modified.)
Most likely, the fixed points (\ref{32}) are just a small subset
of all the fixed points of eq. (\ref{12}) -- the top of an iceberg.
Charge conjugation and translation symmetric states are
not the most general quantum states.
Furthermore, we have concentrated on two-point functions and not
exploited the conserved quantities in higher $n$-point functions
(cf. eq. (\ref{15})). This raises the question what
singles out the thermodynamic fixed point -- an issue which is probably
related closely to the concept of entropy for suitable subsystems.

Finally, one may wonder if the additional fixed points and
conserved correlations appear only in the non-relativistic
approximation (\ref{7}), or if they are also present for relativistic
field equations. The second-order equation (\ref{5}) can always
be written as a coupled set of first order equations\footnote{In
this form they are not of the type (\ref{1}) with hermitean $\hat H$.}
\bear\label{33}
\dot\varphi(x)&=&\pi(x),\quad
\dot\pi(x)=-\frac{\delta S_{cl}}{\delta\varphi^*(x)}\nonumber\\
S_{cl}&=&\int d^dx\{\partial_i\varphi^*(x)\partial_i\varphi(x)
+M^2\rho(x)+\frac{1}{2}\lambda\rho^2(x)\}
\ear
Consider first the particular two-point function\footnote{In
terms of $\chi(x)$ (eq. (\ref{5})) $\tilde\rho_A$ equals
$\rho$ up to corrections $\sim\dot\chi/M$.}
\be\label{34}
\tilde\rho_A(x,y)=\frac{i}{2M}
<\pi(x)\varphi^*(y)-\varphi(x)\pi^*(y)>\ee
It obeys the evolution equation
\bear\label{35}
\partial_t\tilde\rho_A(x,y)&=&-i\{-\frac{1}{2M}(\Delta_x-
\Delta_y)\tilde\rho_A(x,y)\\
&&+\frac{\lambda}{2M}(<\varphi(x)\varphi^*(x)\varphi(x)
\varphi^*(y)>-<\varphi(x)\varphi^*(y)
\varphi(y)\varphi^*(y)>)\}\nonumber\ear
and we immediately recognize the close similarity with
eq. (\ref{24}). Again, if the higher correlation functions
are translation- and charge conjugation-symmetric,
the term $\sim\lambda$ vanishes and $\tilde\rho_A$
evolves according to the von Neumann equation. We emphasize
that in contrast to the non-relativistic equation of motion
charge conjugation symmetry remains conserved for the
relativistic equation. A charge conjugation symmetric initial
state always preserves this property at later time. (In this
context it should be noted that the charge conjugation
$\varphi(p)\to\varphi^*(p),\pi(p)\to\pi^*(p)$ discussed
here is \underline{not} equivalent in
the non-relativistic limit to the transformation $\chi(p)
\to\chi^*(p)$ discussed previously.) In conclusion, as far
as $\tilde\rho_A$ is concerned all initial states with
translation and charge conjugation symmetric $S_0$ are linear
quantum states. For these states $Tr\tilde\rho_A^N$ is
conserved and the system cannot approach the equilibrium fixed
point.

The situation for the relativistic system is, however,
more involved than for the non-relativistic approximation. This
is seen from the time evolution of the other two-point
functions
\bear\label{36}
&&\tilde\rho_\varphi=<\varphi(x)\varphi^*(y)>,\quad
\tilde\rho_\pi=\frac{1}{M^2}<\pi(x)\pi^*(y)>\nonumber\\
&&\tilde\rho_s=\frac{1}{2M}<\pi(x)\varphi^*(y)+\varphi(x)\pi^*(y)>
\ear
which obey
\bear\label{37}
\partial_t\tilde\rho_\varphi(x,y)&=&2M\tilde\rho_s(x,y)\nonumber\\
\partial_t\tilde\rho_s(x,y)&=&M(\tilde\rho_\pi(x,y)-
\tilde\rho_\varphi(x,y))+\frac{1}{2M}(\Delta_x+
\Delta_y)\tilde\rho_\varphi
(x,y)\nonumber\\
&&-\frac{\lambda}{2M}(<\varphi(x)\varphi^*(x)\varphi(x)\varphi^*(y)>
+<\varphi(x)\varphi^*(y)\varphi(y)\varphi^*(y)>)\nonumber\\
\partial_t\tilde\rho_\pi(x,y)&=&-2M\tilde\rho_s(x,y)+\frac{1}{M}
(\Delta_x+\Delta_y)\tilde\rho_s(x,y)\nonumber\\
&&+\frac{i}{M}(\Delta_x-\Delta_y)\tilde\rho_A(x,y)\\
&&-\frac{\lambda}{M^2}(<\varphi(x)\varphi^*(x)\varphi(x)
\pi^*(y)>+<\pi(x)\varphi^*(y)\varphi(y)\varphi^*(y)>)
\nonumber\ear
For $\lambda=0$ the general solution is oscillatory wheras
for nonlinear systems the behaviour of $\tilde\rho_\varphi,
\tilde\rho_\pi$ and $\tilde\rho_s$ is not yet established.
A partial approach to a fixed point is not excluded. In
any case, we observe that for translation-symmetric $\tilde\rho_A
(p,q)=A_2(p)(2\pi)^d\delta(p-q)$ the evolution equations for
$\tilde\rho_\varphi,\tilde\rho_\pi$ and $\tilde\rho_s$
are independent of $\tilde\rho_A$.
Independence of $\tilde\rho_A$ holds trivially also
for the higher correlation functions. Comparing with thermodynamic
equilibrium where $\tilde\rho_A$ vanishes, we obtain again an
infinite set of additional fixed points
\be\label{38}
Z_*^{(A_2)}[j]=Z_*[j,\beta]-iMZ\int\frac{d^dp}{(2\pi)^d} A_2(p)
(j^*_\pi(p)j_\varphi(p)-j^*_\varphi(p)j_\pi(p))\ee
where $j^*_\pi(p)$ and $j^*_\varphi(p)$ are the sources
conjugated to $\pi(p)$ and $\varphi(p)$, respectively.

In conclusion, we find that there is no simple approach
of conservative classical systems to thermodynamic equilibrium.
In many respects the analogy of the evolution of classical
correlation functions with quantum mechanics is striking.
We are all used to the picture that classical mechanics
arises from quantum mechanics
in certain limiting cases. Our findings suggest that the
opposite may also be true: The dynamics of quantum mechanics
and perhaps also quantum statistics follow from a statistical
treatment of classical systems in particular ensembles!

\end{document}